\documentstyle[12pt]{article}
\def\beq{\begin{equation}}
\def\eeq{\end{equation}}
\def\bea{\begin{eqnarray}}
\def\eea{\end{eqnarray}}
\def\bq{\begin{quote}}
\def\eq{\end{quote}}

\def\gappeq{\mathrel{\rlap {\raise.5ex\hbox{$>$}}
{\lower.5ex\hbox{$\sim$}}}}

\def\lappeq{\mathrel{\rlap{\raise.5ex\hbox{$<$}}
{\lower.5ex\hbox{$\sim$}}}}

\begin{document}

\def\be{\begin{equation}}
\def\ee{\end{equation}}
\def\bea{\begin{eqnarray}}
\def\eea{\end{eqnarray}}
\def\lsim{\:\raisebox{-0.5ex}{$\stackrel{\textstyle<}{\sim}$}\:}
\def\gsim{\:\raisebox{-0.5ex}{$\stackrel{\textstyle>}{\sim}$}\:}

\pagestyle{empty}

 \begin{flushright}
 CERN-TH.7481/94\\
 \end{flushright}
\bigskip
\bigskip
\begin{center}
\large {\bf Causal Quantum Mechanics Treating Position and Momentum
Symmetrically} \\
\bigskip
\bigskip
\bigskip
\underline{S.M. Roy}$^\star$ \\
CERN-Geneva, \\
\bigskip
and \\
\bigskip
Virendra Singh \\
Tata Institute of Fundamental Research, Bombay \\
\end{center}
\bigskip
\bigskip
\bigskip

\baselineskip=.80cm
\noindent \underbar{\bf Abstract}: De Broglie and Bohm formulated a
causal quantum mechanics with a phase space density whose integral
over momentum reproduces the position probability density of usual
statistical quantum theory.  We propose a causal quantum theory
with a joint probability distribution such that
the separate probability distributions for position \underbar{and}
momentum agree
with usual quantum theory.  Unlike the Wigner distribution the suggested
distribution is positive definite and obeys the Liouville condition.

\bigskip
\bigskip
\bigskip
\bigskip

\hrule width 6cm
\bigskip

\noindent $^\star$ e-mail: Shasanka@Cernvm.cern.ch
On leave (and address after 4 November 1994)
from Tata Institute of Fundamental Research,
Homi Bhabha Road, Bombay 400 005, India.
\bigskip
\begin{flushleft}
CERN-TH.7481/94\\
\end{flushleft}

\newpage

\pagestyle{plain}
\setcounter{page}{1}

\noindent {\bf 1. \underbar{\bf Introduction}}.  Present quantum theory
does not
make definite prediction of the value of an observable in an individual
observation except in an eigenstate of the observable.  Application of
quantum rules to
two separated systems which interacted in the past together with a local
reality principle (Einstein locality) led Einstein, Podolsky and
Rosen$^1$
to conclude that quantum theory is incomplete.  Bell$^2$ showed that
previous proofs of impossibility of a theory more complete than quantum
mechanics$^3$  made
unreasonable assumptions; he went on however to prove$^4$ that a hidden
variable theory agreeing with the statistical predictions of quantum
theory cannot obey Einstein locality.

Bell's research was influenced by the construction by De Broglie and
Bohm$^5$ (dBB) of a hidden variable theory which reproduced the position
probability density of quantum mechanics but violated Einstein locality
for many particle systems.  For a single particle moving in one
dimension
with Hamiltonian
$$
H = - \hbar^2/(2m) \partial^2/\partial x^2 + U(x),
\eqno (1)
$$
and wave function $\psi (x,t)$, de Broglie-Bohm proposed the complete
description of the state to be $\{\lambda (t),|\psi\rangle\}$, where
$\lambda (t)$ is the instantaneous position of the particle, and its
momentum is
$$
\hat p_{dBB} (\lambda,t) = md\lambda/dt = [{\rm Re}~\psi^\star (-i\hbar
{}~\partial \psi/\partial x)/(|\psi|^2)]_{x=\lambda}.
\eqno (2)
$$
In an ensemble the position density $\rho(\lambda,t)$ agrees with
$|\psi (\lambda,t)|^2$ for all time.  However,
Takabayasi$^6$ pointed out that the joint probability distribution
for position and momentum given by the theory
$$
\rho_{\rm dBB} (\lambda,p,t) = |\psi(\lambda,t)|^2 \delta\left(p - \hat
p_{dBB}(\lambda,t)\right)
\eqno (3)
$$
does not yield the correct quantum mechanical expectation value of $p^n$
for integral $n \neq 1$.  De Broglie$^5$ stated that these values in his
theory ``correspond to the
unobservable probability distribution existing prior to any
measurement'' and measurement will reveal different values distributed
according to standard statistical quantum mechanical formula.  On the
other hand position measurements have no central role since they simply
reveal the existing position.  The asymmetrical treatment of position
and
momentum in the dBB theory constitutes breaking of a fundamental
symmetry
of the quantum theory and has been considered by some physicists as a
defect of the dBB theory (Holland, Ref. 5, p. 21).

Without using hidden variables, Griffiths$^7$ and Gell-Mann and
Hartle$^8$
introduced joint probability distributions for noncommuting observables
at
different times in the consistent history approach to quantum theory of
closed systems.  Wigner$^9$ had earlier introduced a joint distribution
for $x$ and $p$ at the same time,
$$
\rho_W (x,p,t) = \int^\infty_{-\infty} {dy \over 2\pi\hbar} \psi^\star
\left(x + {y \over 2},t\right) \psi\left(x - {y \over 2},t\right)
\exp(ipy/\hbar)
\eqno (4)
$$
which yielded the correct quantum probability distributions separately
for
$x$ and $p$ on integration over $p$ and $x$ respectively.  The Wigner
distribution cannot however be considered a probability distribution
because it is not positive definite, as seen from the fact that the
integral
$$
\int dx~dp~\rho_{W,\psi} (x,p) \rho_{W,\phi} (x,p) =
|(\psi,\phi)|^2/(2\pi\hbar)
$$
vanishes for two orthogonal states $\psi,\phi$.

We wish now to propose a deterministic quantum theory of a closed system
with the following properties.  (We consider in this paper only 1
particle
in 1 space dimension).

\noindent (i) The system point $(x(t),p(t))$ in phase space has a
Hamiltonian flow with
a $c$-number causal Hamiltonian $H_C (x,p,\psi(x,t),t)$ so that in an
ensemble of mental copies of the system the phase space density
$\rho(x,p,t)$ obeys Liouville's theorem
$$
d\rho (x,p,t)/dt = 0.
\eqno (5)
$$
Here $\psi(x,t)$ is the solution of the usual Schr\"odinger equation
$$
i\hbar~\partial \psi(x,t)/\partial t = H\psi(x,t)
\eqno (6)
$$
with $H$ being the standard quantum mechanical Hamiltonian for the
system
and $H_C$ being determined from the following criteria.

\noindent (ii) Each pure ``causal state'', i.e., a set of phase space
points moving according to a single causal Hamiltonian $H_C$ has phase
space density of the deterministic form
$$
\rho(x,p,t) = |\psi(x,t)|^2 \delta\left(p - \hat p(x,t)\right),
\eqno (7)
$$
in which $p - \hat p(x,t) = 0$ not only determines $p$ as a function of
$x$, but also determines $x$ as a function of $p$ at each time (step
functions being allowed when necessary).  Eqn. (7)
guarantees on integration over $p$ the correct quantum probability
distribution in $x$ for any real function $\hat p(x,t)$.  The function
is
determined from the requirement that on integration over $x$,
$\rho(x,p,t)$ should also yield the correct quantum probability
distribution in
$p$.  That such a determination is possible and unique apart from a
discrete 2-fold ambiguity will be a crucial part of the present theory.
It is obvious that our $\hat p(x,t)$ will have to be different from the
$\hat p(x,t)$ of de Broglie-Bohm theory.

\noindent (iii) Since the quantum probability distributions for $x$ and
$p$
in the statistics of many measurements are exactly reproduced, so are
the
standard uncertainty relations.

In Secs. II, III we describe the construction of the momentum $\hat
p(x,t)$ and the causal Hamiltonian $H_C$, in Sec. IV applications to
simple
quantum systems, and in Sec. V conceptual features of the new mechanics.
\bigskip

\noindent {\bf 2. \underbar{\bf Construction of Joint Probability
Distribution of position and}}

\underbar{\bf momentum.}  We seek a positive definite distribution of
the
form (7) where
$\hat p$ is a monotonic function of $x$
$$
\epsilon~\partial\hat p (x,t)/\partial x \geq 0, ~~~~~ \epsilon = \pm 1
\eqno (8)
$$
The monotonicity
property ensures that for a given $t$, the $\delta$-function establishes
one-to-one invertible correspondence between $x$ and $p$ whenever
$\partial \hat p/\partial x$ is finite and non-zero.  (This is the
simplest qualitative assumption about $\hat p (x,t)$ which will be shown
to result in Hamiltonian evolution; in future development we should try
to
replace the monotonicity assumption by the assumption of Hamiltonian
evolution).
The requirement of reproducing the correct quantum probability
distribution of $p$ is that
$$
\int^\infty_{-\infty} \rho(x,p,t)dx = {1 \over \hbar} \big|\tilde
\psi \left({p \over \hbar},t\right)\big|^2,
\eqno (9)
$$
where $\tilde \psi (k,t)$ is the Fourier transform of $\psi(x,t)$.  We
substitute the ansatz (7) into (9) and integrate over momentum to obtain
$$
\int^p_{-\infty} dp' \int_{\hat p (x',t) \leq p} dx' |\psi(x',t)|^2
\delta\left(p' - \hat p(x',t)\right) = \int^p_{-\infty} {dp' \over
\hbar}
\big|\tilde \psi\left({p' \over \hbar},t\right)\big|^2.
\eqno (10)
$$
The region $\hat p(x',t) \leq p$ becomes $x' \leq x$ if $\epsilon = 1$,
and $x' \geq x$ if $\epsilon = -1$, where $\hat p(x,t) = p$.  Thus, we
obtain, for $\epsilon = \pm 1$,
$$
\int^{\epsilon x}_{-\infty} dx' |\psi(\epsilon x',t)|^2 = \int^{\hat
p(x,t)/\hbar}_{-\infty} dk' |\tilde\psi (k',t)|^2.
\eqno (11)
$$
The left-hand side is a monotonic function of $x$ which tends to 1 for
$\epsilon x \rightarrow \infty$ for a normalized wave function; the
right-hand side
is a monotonic function of $\hat p$ tending to 1 for $\hat p \rightarrow
\infty$ (Parseval's theorem).  Hence, for each $t$, Eq. (11)
determines two monotonic functions $\hat p$ of $x$, one for each sign of
$\epsilon$.  (Note that the curve $\hat p (x,t)$ may have segments
parallel
to $x$-axis or $p$-axis corresponding to $\psi(x,t)$ or
$\tilde\psi(p/\hbar,t)$ vanishing in some segment).  The two curves $p =
\hat p_\pm (x,t)$ so determined yield via Eq. (7) phase space densities
$\rho_\pm$, with different causal Hamiltonians $(H_C)_\pm$ determined
below.

\bigskip

\noindent {\bf 3. \underbar{\bf Determination of the Causal
Hamiltonian.}}
We view $\rho(x,p,t)$ as describing an ensemble of
system trajectories in the phase space.  We saw in the last section that
such a description is possible at each time.  We would now like to find
causal Hamiltonian such that the time evolution in phase space implied
thereby is consistent with the time dependent Schr\"odinger equation.

In order that the total number of trajectories is conserved in time we
must have the continuity equation
$$
\partial \rho/\partial t + \partial (\rho
\dot x)/\partial x  + \partial (\rho \dot p)/\partial p  = 0
\eqno (12)
$$
If the dynamics of the trajectories is of Hamiltonian nature i.e.
$$
\dot x = \partial H_C/\partial p, ~~~
\dot p = -\partial H_C/\partial x
\eqno (13)
$$
then we have Liouville's theorem that the phase space density is
conserved,
$$
\partial \rho/\partial t + \dot x \partial \rho/\partial
x + \dot p \partial \rho/\partial p = 0
\eqno (14)
$$
i.e.
$$
\partial \rho/\partial t + (\partial H_C/\partial p)
{}~\partial \rho/\partial x - (\partial H_C/\partial x)
{}~\partial \rho/\partial p = 0.
\eqno (15)
$$
The $c$-number Hamiltonian $H_C$ describing the causal time evolution of
the trajectories in the phase space will be allowed to be different
from the usual $q$-number Hamiltonian $H$ describing the time evolution
of
the Schr\"odinger wave function $\psi$ according to Eq.
(6).

On substituting into Eq. (15) the ansatz (7)
discussed in the last section, we obtain
$$
\xi \delta(p - \hat p) + {\partial \over \partial
p} \left(\eta \delta(p - \hat p)\right) = 0
\eqno (16)
$$
where
$$
\begin{array}{l}
\xi = \partial |\psi|^2/\partial t +
(\partial H_C/\partial p) ~\partial |\psi|^2/\partial x -
\partial \eta/\partial p \\[2mm]
\eta = -|\psi|^2 \left\{\partial \hat p/\partial t +
(\partial \hat p/\partial x) ~\partial
H_C/\partial p + \partial H_C/\partial x\right\}.
\end{array}
$$

We thus need for consistency
$$
\xi = 0 ~~~{\rm and}~~~ \eta = 0 ~{\rm if}~ p = \hat p.
\eqno (17)
$$
We now specialise to the usual case when H is given by Eq. (1).
We find that this situation is taken care of with the choice of $H_C (x,
p, t)$
$$
H_C = {1 \over 2m} (p - A(x,t))^2 + V(x,t).
\eqno (18)
$$
The causal Hamiltonian is of the Newtonian form apart from the
introduction of a vector potential $A(x,t)$ and allowing the potential
$V(x,t)$ to differ from $U(x)$.  Eqs. (17) lead to the following
equations
to determine $V$ and $A$ (after using Schr\"odinger eqn. to substitute
for
$\partial |\psi|^2/\partial t$),
$$
- \partial V(x,t)/\partial x = \partial \hat p(x,t)/
\partial t + (2m)^{-1} \partial\left(\hat p(x,t) -
A(x,t)\right)^2/\partial x,
\eqno (19)
$$
$$
\partial \left[|\psi|^2 (\hat p - A - mv)\right]/\partial x = 0,
\eqno (20)
$$
where $v$ is given by
$$
v(x,t) = \hbar/(2im) ~\partial\ell n (\psi/\psi^\star)/\partial x
\eqno (21)
$$
which is just the de Broglie-Bohm velocity.  Eq. (20) implies that the
quantity in square brackets must be a function of $t$ alone.  We choose
this function of $t$ to be zero in order to avoid a singularity of the
vector potential at the nodes of the wave function.  We thus obtain
$$
A(x,t) = \hat p(x,t) - mv(x,t)
\eqno (22)
$$
With the calculation of the causal Hamiltonian thus completed via Eqs.
(18), (19) and (22) a consistent Liouville description emerges.

It should be stressed that the qualitatively new feature of the theory,
$\hat p(x,t) \neq mv(x,t)$ is quite independent of the specific ansatz
(18) for $H_c$.  Comparison of the continuity equation for the spatial
probability density $\rho(x,t)$ in a deterministic theory with that
following from Schr\"odinger eqn. plus the requirement $\rho(x,t) =
|\psi(x,t)|^2$ imply that $dx/dt = v(x,t)$, the dBB velocity.  The
assumption $\hat p = mv$ will then lead to the dBB answer for momentum
probability density which conflicts with the quantum answer.  Hence, to
reproduce both position and momentum probability densities correctly we
need $\hat p \neq mv$.  On taking ensemble average, the equality is
restored in our theory in agreement with Ehrenfest's theorem.

\bigskip

\noindent {\bf 4. \underbar{\bf Illustrative Examples.}} (i) Quantum
Free
Particle.
Let the quantum free particle be described by the Gaussian momentum
space
wave function
$$
\tilde \psi(p/\hbar,t) = (\alpha\pi)^{-1/4}
\exp\left[-(p-\beta)^2/(2\alpha\hbar^2) - ip^2t/(2m\hbar)\right]
\eqno (23)
$$
so that the coordinate space wave function is
$$
\psi (x,t) =  (\pi \alpha)^{-1/4}
\left(m \alpha/(m + i \alpha \hbar t)\right)^{1/2} \exp
f,
\eqno (24)
$$
$$
f = - (\alpha/2)
\left[(x - \beta t/m)^2 - i\left(\displaystyle{\alpha \hbar t \over m}
x^2
+ \displaystyle{2\beta x \over \alpha \hbar} - \displaystyle{\beta^2 t
\over m \alpha \hbar}\right)\right]\Big/\left(1 + \displaystyle{\alpha^2
\hbar^2 t^2 \over m^2}\right).
$$
Our procedure yields
$$
\hat p - \beta = \pm \hbar \sqrt{{m^2 \alpha^2 \over m^2 + \alpha^2
\hbar^2 t^2}} \left(x - {\beta t \over m}\right),
\eqno (25)
$$
$$
A = (\hat p - \beta) \left[1 \mp {\hbar \alpha t \over \sqrt{m^2 +
(\hbar
\alpha t)^2}}\right],
$$
and
$$
\partial V/\partial x = \pm  (m^2 + \alpha^2\hbar^2
t^2)^{-3/2} [xt(\alpha\hbar)^2 + \beta m] (\hbar\alpha m)
\eqno (26)
$$
The determination of the causal Hamiltonian is now complete apart from
an
irrelevant additive function of $t$.  The quantum potentials $A$ and
$V$ are seen to be proportional to $\hbar$ in this example.  An
interesting feature of Eq. (25) is that
for $\epsilon = +1$, for $t \gg m/(\alpha \hbar) \approx 2(\hbar
E/(\Delta
E)^2)$, it agrees with the naive classical expectation corresponding to
zero vector potential.

\noindent (ii) Quantum Oscillator.  For the minimum uncertainty coherent
state of the harmonic oscillator of mss $m$, angular frequency $\omega$
and
amplitude of oscillation $a$ we find
$$
\rho(x,p,t) = \sqrt{{m\omega \over \pi\hbar}} \exp\left[-{1\over2}
{m\omega \over \hbar} \left(x - a \cos (\omega t)\right)^2\right]
\delta\left(p -\hat p(x,t)\right),
\eqno (27)
$$
where
$$
\hat p(x,t) = -m\omega a \sin(\omega t) \pm m\omega(x-a\cos(\omega t)),
\eqno (28)
$$
and
$$
A(x,t) = \pm m\omega (x - a \cos(\omega t)),
\eqno (29)
$$
$$
- \partial V(x,t)/\partial x = -m\omega^2 a \cos(\omega t) \pm
m\omega^2 a \sin (\omega t)
\eqno (30)
$$
The causal Hamiltonian yields the equation of motion
$$
md^2 x/dt^2 = -m\omega^2 a \cos \omega t
\eqno (31)
$$
which results in exact harmonic motion even for $x$ away from the centre
of the packet.  We do not of course expect this for solutions of the
Schr\"odinger eqn. different from the coherent state here considered.

\bigskip

\noindent {\bf 5. \underbar{\bf New conceptual features.}} (a) We have
derived corresponding to every quantum wave function
$\psi$, two joint probability distributions for position and momentum of
the form (7) which are (i) positive definite, (ii) have Hamiltonian
evolution with causal Hamiltonians $(H_C)_\pm$ and obey (iii)
$$
\int (f(x) + g(p)) \rho_\pm(x,p,t) dx~dp = \left(\psi,\left(f(x) +
g\left(-i\hbar {\partial \over \partial x}\right)\right)\psi\right),
\eqno (32)
$$
for arbitrary functions $f(x)$ and $g(p)$.  Eq. (32) is the major
advantage of the present theory over the $dBB$ theory.  We postpone
the discussion of measurements until we present a generalization of
the theory to many particles.

\noindent (b) Since both $\rho_+$ and $\rho_-$ obey Eq. (32) so will
$\rho
= C\rho_+ + (1-C)\rho_-$ with $0 \leq C \leq 1$.  But since $\rho_+$ and
$\rho_-$ correspond to different causal Hamiltonians $(H_C)_\pm$, $\rho$
will not correspond to a `pure causal state'.  We are led to the concept
of a pure causal state as being more fine grained than a pure wave
function $\psi$.  All $\rho = C\rho_+ + (1-C)\rho_-$ correspond to
$\psi$
$(\rho \leftrightarrow \psi)$ for a continuum of values of $C$, but only
$C = 0,1$ correspond to pure causal states.  To quantum density matrix
states $\Sigma C_\alpha |\psi_\alpha\rangle \langle \psi_\alpha|$
correspond phase space densities $\Sigma C_\alpha \rho_\alpha$ if
$\rho_\alpha \leftrightarrow \psi_\alpha$.

\noindent (c) One can ask if Eq. (32) can be generalized to more general
quantum observables.  Here we face the old problem that there exist
nonclassical observables e.g. $x\left(-i\hbar {\partial \over \partial
x}\right)x$, $\left(\left(-i\hbar {\partial \over \partial x}\right) xx
h.c.\right)/2$ which have different expectation values but the same
`naive' classical analogue $x^2p$.  A trivial but nonunique
way followed already for the $dBB$ distribution is: given a nonclassical
observable $A$ the phase space analogue can be $f(x,p,\psi)$ such that
$f(x,\hat p,\psi) = \psi^\star A\psi/|\psi|^2$.

\noindent (d) It can be proved that for the individual trajectories,
Newton's first law [$d\hat p/dt = 0$ for $U(x) = 0$] holds, unlike
$dBB$ theory.

\noindent (e) A qualitative advantage of our theory over the dBB theory
is
the symmetric treatment of $x$ and $p$ obvious from our phase space
density
$$
\begin{array}{l}
\rho (x,p,t) = |\psi(x,t)|^2 |\tilde\psi(p/\hbar,t)|^2 \hbar^{-1} \delta
\Bigg(\int^{\epsilon x}_{-\infty} dx' |\psi (\epsilon x',t)|^2 \\[2mm]
{}~~~~~~~~~~~~~~~~~~~~~~~~~~~~~~~~~~~~~~~~~~~~~~~~~~ - \int^{\hat
p(x,t)/\hbar}_{-\infty} dk' |\tilde \psi (k',t)|^2\Bigg)
\end{array}
$$

We are deeply indebted to Andr\'e Martin for many discussions and for
a decisive contribution$^{10}$
to the elucidation of conditions that can be
imposed on the phase space density $\rho(\vec x,\vec p,t)$ in higher
dimensions.  We thank
C.V.K. Baba, R. Cowsik, D. Dhar, P. Eberhard, B. d'Espagnat,
J.~Finkelstein, P. Holland, D. Home, S.S. Jha, L. Kadanoff,
A. Kumar, P.
Mitra, G.~Rajasekaran, N.F. Ramsey, D. Sahoo and E.J. Squires for
discussions and correspondence on an earlier version of the work$^{11}$.

\newpage

\noindent {\bf References}

\begin{enumerate}
\item A. Einstein, B. Podolsky and N. Rosen, Phys. Rev. \underbar{47},
777
(1935).

\item J.S. Bell, Rev. Mod. Phys. \underbar{38}, 447 (1966).

\item E.g. J. Von Neumann, ``Mathematische Grundlagen der Quanten
Mechanik'',
Julius Springer Verlag, Berlin (1932) (English Transl.: Princeton Univ.
Press, Princeton, N.J. 1955).

\item J.S. Bell, Physics (Long Island City, N.Y.) \underbar{1}, 195
(1964).

\item L. de Broglie, ``Nonlinear Wave Mechanics, a causal
interpretation'', (Elsevier,
Amsterdam 1960); D. Bohm, Phys. Rev. \underbar{85}, 166, 180 (1952); D.
Bohm, B.J. Hiley and P.N. Kaloyerou, Phys. Rep. \underbar{144}, 349
(1987);  D. D\"urr, S. Goldstein and N. Zanghi, J. Stat. Phys.
\underbar{67}, 843 (1992);
P.R. Holland,  ``The Quantum Theory of Motion''
(Cambridge Univ. Press, Cambridge 1993).

\item T. Takabayasi, Progr. Theoret. Phys. \underbar{8}, 143 (1952).

\item R.B. Griffiths, J. Stat. Phys. \underbar{36}, 219 (1984); Phys.
Rev.
Letts. \underbar{70}, 2201 (1993).

\item M. Gell-Mann and J.B. Hartle, in Proc. 25th Int. Conf. on High
Energy Physics, Singapore, 1990, Eds. K.K. Phua and Y. Yamaguchi (World
Scientific, Singapore 1991).  R. Omn\`es, Rev. Mod. Phys. \underbar{64},
339 (1992).

\item E. Wigner, Phys. Rev. \underbar{40}, 749 (1932).

\item A. Martin and S.M. Roy (in preparation).

\item S.M. Roy and V. Singh, `Deterministic Quantum Mechanics in One
Dimension', to appear in the Proceedings of the Conference ``Bose and
Twentieth Century Physics'', Calcutta, January 1994
(Kluwer Academic, Dordrecht, Netherlands).

\end{enumerate}

\end{document}